\documentclass[]{aa}
\usepackage{graphicx}
\usepackage{amssymb}
\begin{document}
   \title{1RXSJ062518.2+733433: A bright, soft intermediate polar\thanks{Partly based on observations collected at the Wendelstein observatory, 
operated by the Universit{\"a}ts-Sternwarte M{\"u}nchen.}}

   \author{A. Staude
          \inst{1},
	   A.D. Schwope
	  \inst{1},
          M. Krumpe
	   \inst{1},
	   V. Hambaryan
	  \inst{1},
          \and
	   R. Schwarz
	  \inst{2}
          }

   \offprints{astaude@aip.de}

   \institute{Astrophysikalisches Institut Potsdam (AIP),
              An der Sternwarte 16, D-14482 Potsdam, Germany\\
         \and
             Universit{\"a}tssternwarte G{\"o}ttingen, 
	     Geismarlandstra{\ss}e 11, D-37083 G{\"o}ttingen, Germany\\
             }

   \date{Received; accepted}

   \abstract{We present the results of 50 hours time-resolved
   $R$-band photometry of the ROSAT all-sky survey source
   1RXSJ062518.2+733433. The source was identified by Wei et
   al.~(\cite{wei}) as a cataclysmic variable. Our photometry, performed
   in 10 nights between February 11, 2003, and March 21, 2003, reveals
   two stable periodicities at 19.7874 and 283.118 min, which are identified as
   probable spin and orbital periods of the binary. We therefore
   classify 1RXSJ062518.2+733433 as an intermediate polar.
   Analysis of the RASS X-ray observations reveal
   a variability of 100\% in
   the X-ray flux and a likely soft X-ray excess. The new IP thus 
   joins the rare group of soft IPs with only four members so
   far.
   }

   \authorrunning{A. Staude et al.}
   \titlerunning{1RXSJ062518.2+733433: A bright, soft intermediate polar}

   \maketitle
%

\section{Introduction}

\object{1RXSJ062518.2+733433} (RX06 in the following) was discovered
during an AGN search in the ROSAT All-Sky Survey (RASS) by Wei et
al.~(\cite{wei}) and  identified as a cataclysmic variable star
(CV). Wei et al.~selected their sources for an optical identification
program according to a high ratio of X-ray to optical flux 
($F_{X}/F_{opt}$). Most of their objects are indeed
emission line AGNs, however, four CVs are among them. 
The optical spectrum of RX06 showed a blue continuum with
bright emission lines of hydrogen and helium, both neutral and
ionized, thus making the object a candidate magnetic CV. 

New magnetic CVs were found numerously in the RASS, most of them are
polars (AM Herculis stars). The strong magnetic fields in these
systems channel the accreted matter to rather small regions at the
footpoints of accreting field lines, where a soft X-ray engine is
located. As a result of the RASS and several identification programs
the total number of polars has increased from a dozen or so
pre-ROSAT ones to about 70 now. Progress for the cousins with weaker
magnetic fields, the DQ Herculis stars or intermediate polars (IPs),
was much slower. Today only about 21 confirmed systems are
known. This is mainly due to their relative hard X-ray spectra,
preventing their detection in large number in the RASS or the EUVE all
sky survey. 

IPs were traditionally thought to have accretion disks truncated by
the magnetic field in the inner regions. They accrete from the disk
via an azimuthally extended accretion curtain.
Although the processes of energy release at
the foot points of the accretion curtain are similar to polars, the
classical IPs lack detection of a soft X-ray component.
This was usually explained by a too low temperature of the soft
component or by high photoelectric absorption in the accretion
curtain. ROSAT has discovered a small number of weakly absorbed, so
called soft IPs (Mason et al.~\cite{mason}, Haberl et al.~\cite{ha-tho}, 
\cite{ha-mo}) with
relatively high magnetic field strength and corresponding high
collimation of the accretion flow. Here we present the discovery of a
new, soft IP. The classification rests on extended optical photometry
and an analysis of the RASS X-ray data.

\section{Optical photometry}

Our photometric observations were performed with the 0.7m-telescope of
the AIP in Potsdam and the 0.8m-telescope of the Universit{\"a}t
M{\"u}nchen at the Wendelstein in the period between February 11, 2003
and March 21, 2003. The Potsdam telescope is equipped with
a cryogenic Photometrics CCD, using a thinned SITe 1k x 1k chip. The 
telescope at Mount Wendelstein was operated with the MONICA CCD camera system 
(Roth \cite{roth}).  
All observations were performed using a broad-band $R$ filter.
Details are shown in Table~\ref{obslog}.  

\begin{table}
\caption[]{Log of observations.}
\label{obslog}

\begin{tabular}{llcrc}
\noalign{\smallskip} \hline \noalign{\smallskip}
Date&Telescope&Filter&Duration&Exp. time\\
\noalign{\smallskip} \hline \noalign{\smallskip}
2002/02/11&AIP 0.7m&$R$&11.34 h&60 s\\
2002/02/24&AIP 0.7m&$R$&4.42 h&60 s\\
2002/02/25&AIP 0.7m&$R$&5.67 h&60 s\\
2002/02/26&AIP 0.7m&$R$&4.67 h&60 s\\
2002/03/16&UM 0.8m&$R$&3.02 h&35 s\\
2002/03/17&UM 0.8m&$R$&4.09 h&35 s\\
2002/03/18&UM 0.8m&$R$&4.90 h&35 s / 25 s\\
2002/03/19&UM 0.8m&$R$&4.50 h&65 s / 35 s\\
2002/03/20&UM 0.8m&$R$&2.13 h&35 s\\
2002/03/21&UM 0.8m&$R$&5.03 h&35 s\\
\noalign{\smallskip} \hline \noalign{\smallskip}
\end{tabular}
\end{table} 

\begin{figure}
\fbox{\resizebox{\hsize}{!}{\includegraphics[clip]{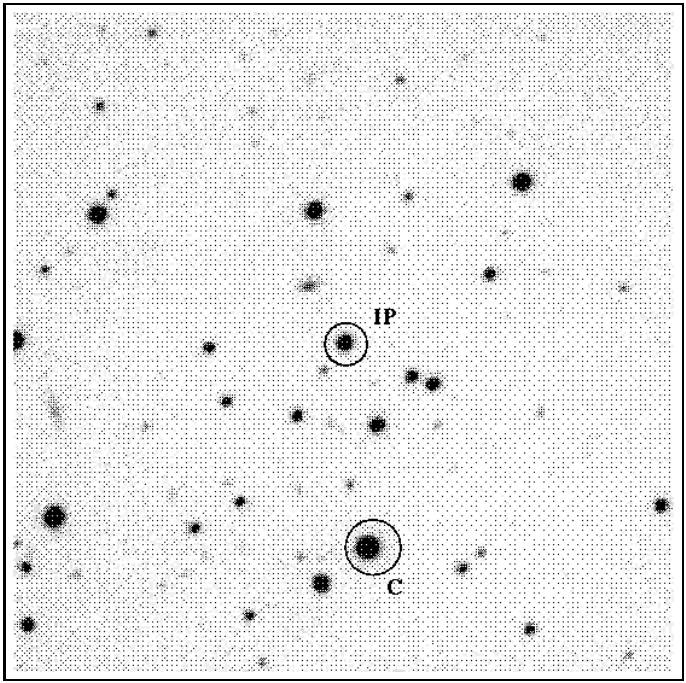}}}
\caption{$R$-band image of RX06 taken with the AIP 0.7m-telescope ($\sim
6' \times 6'$). North is on top and East is left. The object and the
comparison-star ('C') are marked.} 
\label{finder}
\end{figure}

The brightness of the object was measured with respect to a brighter
comparison star (marked 'C' in Fig.~\ref{finder}), for which we
determined an apparent $R$-magnitude of $12\fm89 $ from standard-star
measurements on February 24 and March 18, 2003. 

\begin{figure}
\resizebox{\hsize}{!}{
\includegraphics[clip]{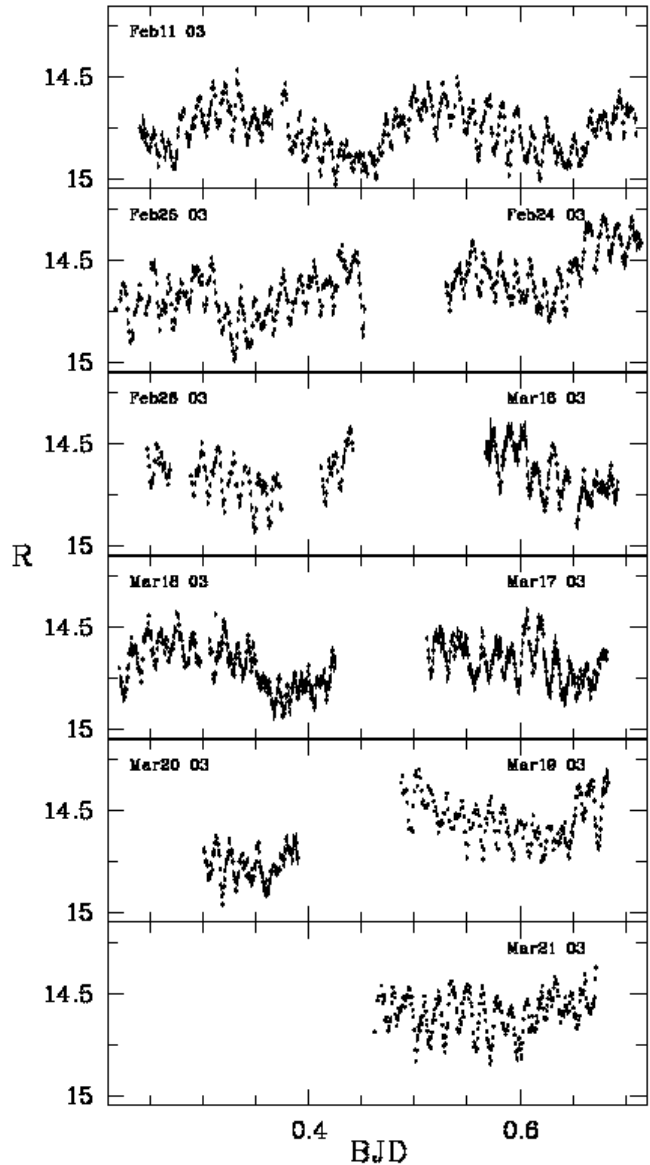}}
\caption{All light curves used for the period determination in this
paper. The brightness is given in $R$-magnitudes, time in fractions of
barycentric JD. The light curves of Feb 25, March 18 and 20 have been 
offset horizontally by $-0.25$ days.} 
\label{allobs}
\end{figure}

Fig.~\ref{allobs} shows all our observations. Obviously there is the
superposition of brightness variations on two different time-scales: 
short-term oscillations with a period of $\sim$ 20 min and broad humps
with a periodicity of $\sim$ 5 hours. In addition the overall brightness
level does vary on a night-to-night basis by about 0.2~mag.

\section{Period analysis}

\begin{table}
\caption[]{Timings of the pulse maxima.}
\label{pulsemax}
\begin{center}
\begin{tabular}{lrrrc}
\noalign{\smallskip} \hline \noalign{\smallskip}
\multicolumn{1}{c}{$T_{max}$ } & \multicolumn{1}{c}{Cycle}  & OMC \\ 
\multicolumn{1}{c}{(BJD 2450000+)} &  & (sec)   \\ 
\noalign{\smallskip} \hline \noalign{\smallskip}
2682.41805 & $   0 $ & $-11$ \\
2695.55464 & $  956$ & $-17$ \\
2696.51701 & $ 1026$ & $ 23$ \\
2697.34144 & $ 1086$ & $ 19$ \\
2715.63071 & $ 2417$ & $-12$ \\
2716.56483 & $ 2485$ & $-37$ \\
2717.52629 & $ 2555$ & $-74$ \\
2718.55801 & $ 2630$ & $ 22$ \\
2719.57546 & $ 2704$ & $ 74$ \\
2720.56402 & $ 2776$ & $  4$ \\
\noalign{\smallskip} \hline \noalign{\smallskip}
 \end{tabular}
\end{center}
 \end{table}


\begin{figure*}
 \includegraphics[angle=-90,width=\textwidth,clip]{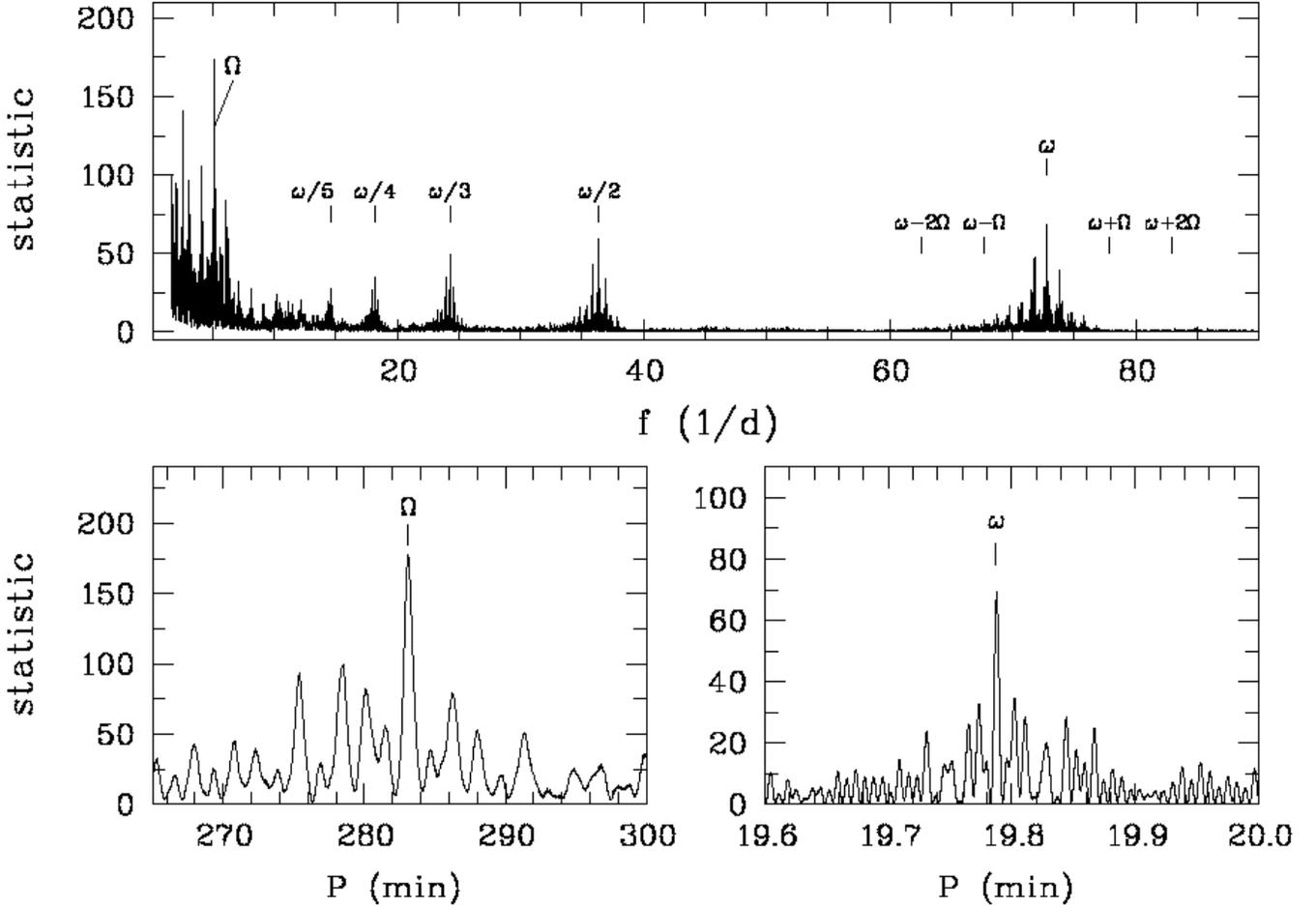}
\caption{The results of the period analysis. {\em top}: full-range periodogram,  {\em left}: around 283 min, {\em right}: around 19.8 min,}
\label{periodogram}
\end{figure*}

\begin{figure}
\resizebox{\hsize}{!}{
\includegraphics[clip]{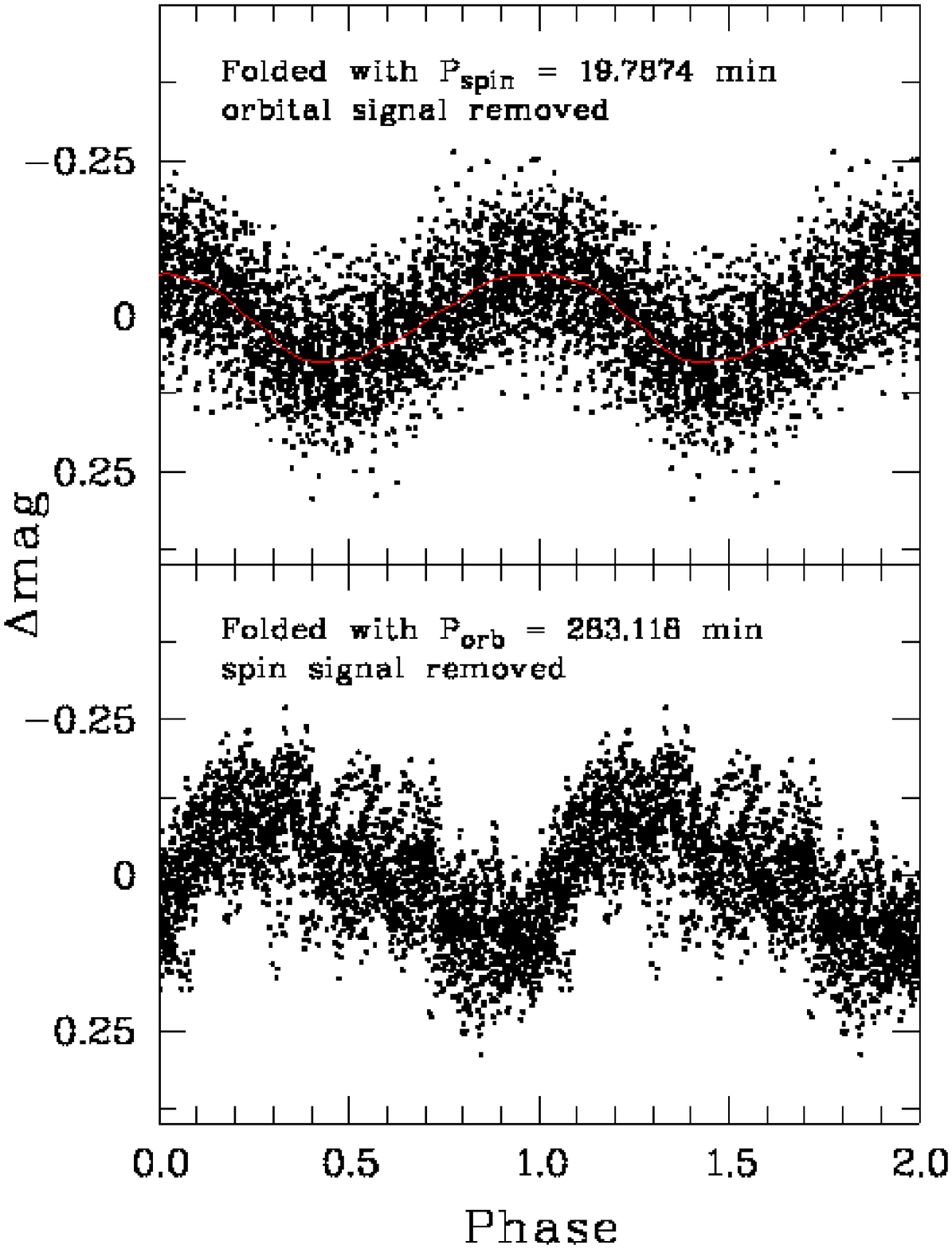}}
\caption{2865 photometric measurements folded with the spin (upper
panel) and orbital ephemeris (lower panel) given by Eqs.~1 and 2.
The data have been normalised and the competing signal (e.g. 
the orbital signal in case of the spin folded data) has been
subtracted. Two cycles shown for clarity.} 
\label{lc_fold}
\end{figure}

For determining the periodicities we applied the analysis-of variance
method (AoV, Schwarzenberg-Cerny \cite{schw-c}) on the combined data
set including all nights. Beforehand the individual data sets were 
normalised to the mean brightness level, in order to account for the 
night-to-night variations.

The resulting periodogram (Fig.~\ref{periodogram}) is dominated by
two most significant periods at 19.7877 min (72.774 cyc/d) min and 
283.117 min (5.086 cycle/d), which 
we interprete as spin and orbital period of a DQ Herculis system, 
respectively. 
Other significant peaks in the periodogram include the inevitable 
alias periods due to sampling, a series of sub-harmonics of the spin period 
as well as the first sub-harmonic of the orbital signal. 
In the raw periodogram any of the side-band periods
commonly dectected in the intermediate polars are hidden in the
complicated alias structure. When repeating the analysis 
after removal of the two main signals from the light curves, a clear signal at
21.2737 min becomes evident. If our assignment of spin and orbital period 
were correct, this detection would represent the $\omega-\Omega$ side-band
frequency.

The accuracy of the spin period (FWHM of the peak) is   
sufficient to establish an 
unequivocal cycle count in our data.
Individual barycentric timings of the pulse maxima for each night of 
observation were determined from fitting a sinusoid to our smoothed 
light curves and are listed in Table~\ref{pulsemax}.
They obey a linear relation yielding a spin ephemeris
\begin{equation}
\label{e:spin}
{\rm BJD(max)} = 245\, 2682.4181(5) + E \times 0.01374128(16)
\end{equation}
where the number in brackets denote the uncertainties 
in the last digits as given by the linear regression.
The scatter of the measured values is in general in agreement with 
the formal errors of our sine-fitting.

In case of the orbital modulation the sharp rise from minimum provided 
the best marker for defining the zero point of our orbital ephemeris, 
\begin{equation}
\label{e:orb}
 {\rm BJD(orb)} = 245\, 2682.464(3) + E \times 0.19661(27)
\end{equation}
while the period and its uncertainty were taken from the peak 
value in the periodogram. 
Other orbital minima, possibly useful for defining a future
long-baseline ephemeris, were measured at barycentric Julian dates  
2452682.65, 2452695.633, 2452696.58, 2452717.619, 2452718.644 and
2452719.614. Given the irregular shape of the minima these timings
are not accurate to better than 5 min.

In order to determine the waveform of the spin and orbital modulations,
we normalised all available data to the nightly mean and subtracted 
the competing signal, e.g. the mean orbital variation in case of the
spin pulse. The resulting data folded over spin and orbital period given
by Eqs.~1 and 2 are shown in Fig.~\ref{lc_fold}. The spin pulse
is almost a sine-wave and has a semi-amplitude of 0.07 mag, comparable
to other DQ Her stars with similar spin periods. In principle 
the actual spin period may also be twice the observed value, and the pulse 
should then be double peaked, with unequal maxima or minima as seen in
the few double pulsed IPs (e.g. V405 Aur, Skillman \cite{skill}). 
To check this possibility the data were also folded 
using twice the spin period (39.57 min). Contrary to the known double
pulsed IPs there are no evident asymmetries
between the two pulses and we conclude that 19.7874 min is the true spin
period of the white dwarf in RX06.

Another supporting argument that our spin period is
correct comes from the probable detection of the beat frequency at
$(\omega-\Omega)$. In contrast, a double-peaked pulse would imply that 
this signal corresponds to a frequency at $(\omega/2)-\Omega$, which has 
not been observed in IPs yet.

The orbital wave form is quite asymmetric with a sharp rise and 
smoother decline, similar to the behaviour of BG CMi (Patterson \&
Thomas \cite{patt93}). 
Such orbital variations in intermediate polars are in general thought 
to be the result of X-ray heating of either the secondary star, or/and 
the bright spot at disk rim. In order to observe such an effect with a
peak-to-bottom amplitude of 0.2~mag, the inclination of RX06 is probably
not very low, i.e. $\gtrsim30^{o}$.


\begin{figure}
\resizebox{\hsize}{!}{\includegraphics[clip=true]{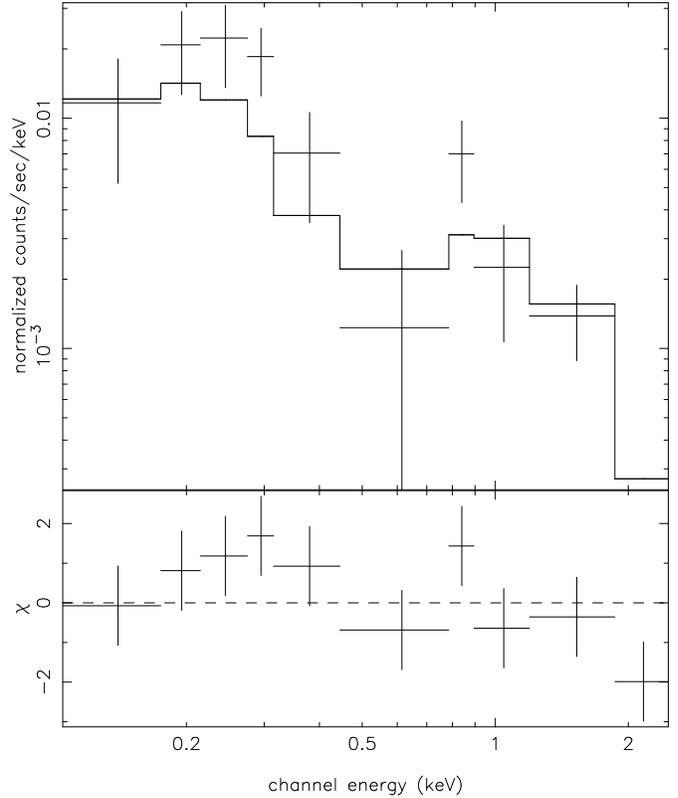}}
\caption{Spectrum of the RASS X-ray data. They were fitted with a
single bremstrahlung component of temperature corresponding to
$kT_{\rm tb} = 15$\,keV. The residuals show excess emission in the
soft band between 0.2 and 0.4 keV.}
\label{rass_spec}
\end{figure}

\section{ROSAT all-sky survey observations}
RX06 was scanned during the RASS 29 times and got a total exposure of
462\,s. The ROSAT Bright Source Catalogue (Voges et al.~\cite{voges}) lists
the source with a mean count rate of 0.18 s$^{-1}$, and hardness
ratios HR1 $= -0.32 \pm 0.11$ and HR2 $= -0.02 \pm 0.20$. X-ray photons
of the field of RX06 were extracted from the ROSAT data archive at
MPE. A light curve of RX06, mean count rate per individual scan, was
generated by subtracting the background signal measured in
concentric circles from the source photons and dividing by the duration of
individual scans. This light curve shows 100\% variability of the
X-ray source with maximum count rate of 0.6\,s$^{-1}$. Folding of
the light curve over both, the spin and the orbital periods, does not
yield an obvious periodic pattern of variability, clearly due to the
short exposure. 

The observed hardness ratio HR1 indicates a rather soft X-ray
spectrum. An X-ray spectrum of RX06 was built from the 83 source
photons detected in the RASS. It was fitted with a hard
bremstrahlung model ($kT_{\rm tb} = 15$\,keV fixed, plus interstellar cold
absorption), with a pure black-body model ($kT_{\rm bb}$ free, plus
interstellar absorption), and with a combined black-body/bremsstrahlung
model (bb/tb). The absorbed black-body could not represent the spectrum at all
and is clearly ruled out. The 15 keV bremsstrahlung spectrum with zero 
interstellar absorption gives a marginal good fit to the data. 
The reduced $\chi^2$ of the fit is 1.6 for 8 degrees of freedom ($dof$) and
the null hypothesis probability is 12\%. The fit and the residuals are
shown in Fig.~\ref{rass_spec} and indicate an excess of soft photons
between 0.2 and 0.4 keV. Taking into account some absorption, the soft
excess increases. This makes a combined bb/tb model very likely. 
Such a fit with $kT_{\rm tb} = 15$\,keV (fix) gives $N_H =
(2.8\pm0.9)\times 10^{20}$\,cm$^{-2}$, 
$kT_{\rm bb} = 43 \pm 9$\,eV, at reduced $\chi^2 = 1.00$ for 6 $dof$.
We regard the presence of a soft component as very likely.

\begin{figure}
\resizebox{\hsize}{!}{\includegraphics[clip=true]{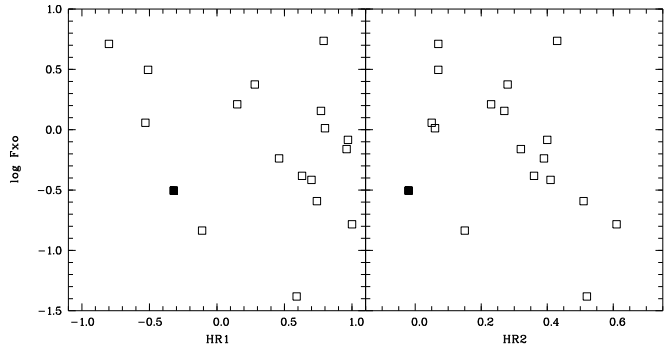}}
\caption{Logarithm of the X-ray to optical flux ratio as a function of
RASS hardness ratios of all confirmed IPs. The new IP RX06 is
indicated by the filled square. }
\label{fxo_hrs}
\end{figure}

\section{Discussion and conclusions}

RX06 is likely to be a cataclysmic variable of DQ Her-type, an
intermediate polar. It exhibts the defining criterion of these systems: two
stable and coherent periodicities in the optical photometry of 19.7874\,min and
283.118\,min which we tentatively regard as the spin and orbital
periods of the binary. The orbital and spin modulation  of RX06 closely
resemble those of other well-known intermediate polars like e.g.~BG CMi
(e.g. Patterson \& Thomas~\cite{patt93}). 
The absence of any obvious eclipse-like feature limits the orbital
inclination to less than 75 degrees.

With our measured spin period of $P_{spin} \simeq 0.1 P_{\rm orb}$ and 
an orbital period above the period gap, the system has parameters similar 
to most previously known classical DQ Her stars 
(see Patterson \cite{patterson}, Hellier \cite{hellier-keele}, 
\cite{hellier-anna}, \cite{hellier-cvs}). 
Other hallmarks of magnetic CVs observed in RX06 include
an optical
spectrum with a continuum steeply increasing towards the blue with H,
HeI and HeII emission lines superposed, X-ray
variability by 100\%, and an X-ray spectrum likely consisting of a hard
(bremsstrahlung-like) and a soft (bb-like) component. 
The optical emission likely originates 
from the hot accretion disk and the
magnetically channelled accretion stream or curtain.
The presence of the HeII line is a clear
additional indication of a soft X-ray component in the system.


In Fig.~\ref{fxo_hrs} we put RX06 in context with other confirmed IPs.
These were drawn from the compilation of K.~Mukai
(http://lheawww.gsfc.nasa.gov
/users/mukai/iphome/iphome.html). In the
figure we plot the ratio of X-ray to optical flux as a function
of the RASS hardness ratios HR1 and HR2. 
Five out of 23 IPs were not discovered in the RASS. 
The X-ray and optical fluxes were computed using the relations 
$F_{\rm X} = 10^{-11} \times CR_{\rm RASS}$ and $F_{\rm opt}
= 10^{(-0.4V - 5.44)}$. 
The diagram has limited analytic power, since optical and X-ray data
were not obtained simultaneously. 
However, compared with the other IPs, RX06 has a rather
low flux ratio $F_{\rm X}/F_{\rm opt}$, and a very soft X-ray
spectrum. The three systems with smaller value of HR1 than that of
RX06 (PQ Gem, V405 Aur = 0558+53, and UU Col = 0512--32), are well
established soft IPs, i.e.~systems with a hard plasma cooling
component of typical temperature $10-30$\,keV, and a soft bb-like
component similar to the spectra of the strongly magnetic 
AM Herculis binaries. RX06 has the smallest HR2 of all
the IPs, which supports the classification as soft IP. 

The presently available data are too sparse in order to discern
between the possible accretion modes, stream-fed, disk-fed or a
mixture of both. There is also the possibility that the short-term 
variations do not trace
the spin but a side-band period, as seen for example in V1223\,Sgr 
(Warner \cite{warner}). Both issues should be clarified by a
dedicated X-ray observation of RX06. 
\begin{acknowledgements}
We would like to thank H. Barwig  for the
generous allocation of observing time at the Wendelstein observatory,
and O. B\"arnbantner and C. Ries for assitance with these observations.
RS acknowledges the support by the DLR under 
grant \mbox{50 OR 0206}. 
Many thanks to the referee of this paper, A.J. Norton, for his helpful 
comments.
We have made use of the ROSAT Data Archive of the Max-Planck-Institut 
f{\"u}r extraterrestrische Physik (MPE) at Garching, Germany.
\end{acknowledgements}


\begin{thebibliography}{}

    \bibitem[1994]{ha-tho} Haberl, F., Thorstensen, J., Motch, C., et al.~1994, 
      A\&A, 291, 171

    \bibitem[2002]{ha-mo} Haberl, F., Motch, C., Zickgraf, F.-J.~2002, 
      A\&A, 387, 201

   \bibitem[1996]{hellier-keele} Hellier, C. 1996,
      in Proc. of 158$^{th}$ IAU colloquium, Kluwer Academic Publishers, Dordrecht

   \bibitem[1999]{hellier-anna} Hellier, C. 1999,
      in ASP Conf. Ser. Vol 157, 1

   \bibitem[2001]{hellier-cvs} Hellier, C. 2001,
      Cataclysmic Variable Stars, Springer-Verlag UK

   \bibitem[1992]{mason} Mason, K., Watson, M., Ponman, T., et al.~1992, 
      MNRAS, 258, 749

   \bibitem[1993]{patt93} Patterson, J., Thomas, G.~1993,
      PASP, 105, 59

   \bibitem[1994]{patterson} Patterson, J.~1994,
      PASP, 106, 209


   \bibitem[1990]{roth} Roth, M. M. 1990,      
      in ASP Conf. Ser. Vol. 8, 380

   \bibitem[1996]{skill} Skillman, D. R.~1996,
      PASP, 108, 130

   \bibitem[1989]{schw-c} Schwarzenberg-Cerny, A. 1989, 
      MNRAS, 241, 153

   \bibitem[1999]{voges} Voges, W., Aschenbach, B.,  Boller, T. et al.~1999, 
      A\&A, 349, 389

   \bibitem[1986]{warner} Warner, B.~1986,
      MNRAS, 219, 347

   \bibitem[1999]{wei} Wei, J. Y., Xu, D. W., Dong, X. Y. \& Hu, J. Y. 1999,
      A\&AS, 139, 575



\end{thebibliography}
\end{document}